\documentstyle[aas2pp4]{article}  
\lefthead{Fassnacht et al.}
\righthead{}
\begin{document}

\title{B2045+265: A New Four-Image Gravitational Lens from 
CLASS\footnote{Based on observations made with the National Radio
Astronomy Observatory, which is operated by Associated Universities,
Inc., under cooperative agreement with the National Science
Foundation; with the NASA/ESA {\em Hubble Space Telescope}, obtained
at the Space Telescope Science Institute which is operated by AURA,
Inc.\ under NASA contract NAS 5-26555; and with the W. M. Keck
Observatory, which is operated as a scientific partnership among the
California Institute of Technology, the University of California, and
the National Aeronautics and Space Administration.  The Keck
Observatory was made possible by the generous financial support of the
W.M. Keck Foundation.}}

\author{\sc 
	C.\ D.\ Fassnacht, 
	R.\ D.\ Blandford\altaffilmark{2}, 
	J.\ G.\ Cohen,
	K.\ Matthews, 
	T.\ J.\ Pearson, 
	A.\ C.\ S.\ Readhead, 
	D. S.\ Womble\altaffilmark{3}
}
\affil{
	Palomar Observatory, 105-24, 
	California Institute of Technology, 
	Pasadena, CA 91125 \\
}

\author{\sc 
	S.\ T.\ Myers
}
\affil{
	Department of Physics and Astronomy, 
	University of Pennsylvania, 
	209 S. 33rd St., 
	Philadelphia, PA 19104-6396
	\\
}

\author{\sc 
	I.\ W.\ A.\ Browne, 
	N.\ J.\ Jackson, 
	D.\ R.\ Marlow, 
	P.\ N.\ Wilkinson
}
\affil{
	NRAL Jodrell Bank,
	University of Manchester, 
	Macclesfield, 
	Cheshire SK11 9DL, UK \\
}

\author{\sc 
	L.\ V.\ E.\ Koopmans
} 
\affil{
	Kapteyn Astronomical Institute,
	Postbus 800,
	9700 AV Groningen,
	The Netherlands \\
}

\author{\sc 
	A.\ G.\ de Bruyn\altaffilmark{4},
	R.\ T.\ Schilizzi,
}
\affil{
	NFRA, 
	Postbus 2,
	7990 AA Dwingeloo,
	The Netherlands \\
}

\author{\sc 
	M.\ Bremer, 
	G.\ Miley 
}
\affil{
	Leiden Observatory,
	Postbus 9513,
	2300 RA Leiden,
	The Netherlands \\
}
\altaffiltext{2}{
Also Theoretical Astrophysics, 130-33, 
     California Institute of Technology, 
     Pasadena, CA 91125
}
\altaffiltext{3}{
Present address: Monterey Institute for Research in Astronomy,
200 Eighth St., Marina, CA 93955, dsw@mira.org
}
\altaffiltext{4}{
Also Kapteyn Astronomical Institute,
	Postbus 800,
	9700 AV Groningen,
	The Netherlands
}

\begin{abstract}

We have discovered a new gravitational lens in the Cosmic Lens All-Sky
Survey (CLASS).  The lens B2045+265 is a four-image system with a
maximum separation of 1\farcs9.  A fifth radio component is detected,
but its radio spectrum and its positional coincidence with infrared
emission from the lensing galaxy strongly suggests that it is the
radio core of the lensing galaxy.  This implies that the B2045+265
lens system consists of a flat-spectrum radio source being lensed by
another flat-spectrum radio source.  Infrared images taken with the
{\em Hubble Space Telescope} and the Keck I Telescope detect the
lensed images of the background source and the lensing galaxy.  The
lensed images have relative positions and flux densities that are
consistent with those seen at radio wavelengths.  The lensing galaxy
has magnitudes of $J = 19.2^m$, $m_{F160W} = 18.8^m$ and $K = 17.6^m$
in a 1\farcs9 diameter aperture, which corresponds to the size of the
Einstein ring of the lens.  Spectra of the system taken with the Keck
I Telescope reveal a lens redshift of $z_{\ell} = 0.8673$ and a source
redshift of $z_s = 1.28$.  The lens spectrum is typical of a Sa
galaxy.  The image splitting and system redshifts imply that the
projected mass inside the Einstein radius of the lensing galaxy
is $M_E = 4.7 \times 10^{11} h^{-1} M_{\sun}$ .  An estimate of the
light emitted inside the Einstein radius from the $K$ magnitude gives
a mass-to-light ratio in the rest frame $B$\ band of $(M/L_B)_E = 20\,
h (M/L_B)_{\sun}$.  Both the mass and mass-to-light ratio are higher
than what is seen in nearby Sa galaxies.  In fact, the implied
rotation velocity for the lensing galaxy is two to three times higher
than what is seen in nearby spirals.  The large projected mass inside
the Einstein ring radius may be the result of a significant amount of
dark matter in the system, perhaps from a compact group of galaxies
associated with the primary lensing galaxy; however, it may also arise
from a misidentification of the source redshift.  A simple model of
the gravitational potential of the lens reproduces the image positions
well, but further modeling is required to satisfy the constraints from
the image flux density ratios.  With further observations and modeling,
this lens may yield an estimate of $H_0$.

\end{abstract}


\section{Introduction}

The Cosmic Lens All-Sky Survey (CLASS; \cite{class}) is a large-scale
survey for gravitational lenses among flat-spectrum radio sources.
The primary goals of CLASS are to find lenses which may be suitable
for determinations of the Hubble Constant, $H_0$ (e.g., \cite{refsdal};
\cite{bn}), and to study the lensing rate in a large, homogeneous
sample in order to place limits on the cosmological constant,
$\Lambda$ (e.g., \cite{tog}; \cite{turner}; \cite{ffk}; \cite{ft}).
Over 12,000 sources have been observed with the VLA in three sessions
-- the first in the spring of 1994 (CLASS~1; $\sim$3300 sources), the
second in the summer of 1995 (CLASS~2; $\sim$4500 sources) and the
third in the spring of 1998 (CLASS~3; $\sim$5000 sources).  The vast
majority of flat-spectrum radio sources are dominated by emission from
a single compact core; all CLASS sources containing multiple compact
components are selected as gravitational lens candidates, amounting to
50 -- 100 candidates in each phase of the survey.  These are followed
up with high resolution radio imaging using MERLIN; the best surviving
candidates are then imaged using the VLBA.  During this procedure, the
majority of VLA candidates are rejected based on surface brightness
and morphology criteria.  The candidates that survive the radio
filters are then investigated further with optical and/or infrared imaging,
spectroscopy and more radio imaging.  By these means we have
discovered 11 new lenses in the survey, in addition to the one
reported here.   We are investigating an
additional $\sim$25 promising lens candidates.

We use $H_0 = 100 h\ {\rm km\ s}^{-1}\ {\rm Mpc}^{-1}$ and,
except where noted, assume $q_0 = 0.5$\ throughout this paper.

\section{VLA Observations}

The lens system B2045+265 (GB6 J2047+2643) was observed on 1995
September 02 as part of CLASS~2.  The observations were made at
8.4~GHz with the VLA in A configuration, giving a resolution of
$\sim$0\farcs25.  Observation and data reduction techniques for the
survey will be discussed in Myers et al.\ (1998).  The survey image of
B2045+265 shows four components, with a possible detection of a fifth,
weak component.  The system has a standard lens geometry
(e.g., \cite{bn}) and is similar in appearance to B1422+231
(\cite{p1422}).

The source was re-observed with the VLA on 1996 December 31.
Observations were made in A configuration at 1.4, 4.9, 8.5 and
14.9~GHz.  Details of the observations are given in 
Table~{\ref{tab_af311obs}}.  The source 3C\,286 was used as a flux
calibrator.  Phase calibrators were selected from the VLA
calibrator list.  The data were calibrated using standard AIPS
routines, and maps were made using DIFMAP (\cite{difmap}).  At the
frequencies above 1.4~GHz, the source can be characterized as five
point components, with no significant extended structure.  Hence, the
maps were made by fitting five point components to the data and then
repeating a cycle of model-fitting and phase self-calibration.  The
self-calibration time scale was set to the length of the individual
scans at each frequency.  Both the positions and the flux densities were
allowed to vary in the model fitting.  At 1.4~GHz the beam size is
large compared to the component separation, so the model fitting was
done by fixing the component positions at their 8.5~GHz values and
varying only the component flux densities.  In addition, there were
confusing sources in the 1.4~GHz map that had to be included as
components in the model.

\begin{deluxetable}{llrrl}
\tablewidth{0pt}
\scriptsize
\tablecaption{Radio Observations\label{tab_af311obs}}
\tablehead{
\colhead{}
 & \colhead{}
 & \colhead{$\nu$}
 & \colhead{$t_{tot}$}
 & \colhead{Angular}
 \\
   \colhead{Array}
 & \colhead{Date}
 & \colhead{(GHz)}
 & \colhead{(min)}
 & \colhead{Resolution}
}
\startdata
VLA    & 1995 Sep 02 &  8.5 &  0.5 & 0\farcs25 \\
VLA    & 1995 Sep 11 & 14.9 &   7  & 0\farcs14 \\
VLBA   & 1995 Nov 12 &  5.0 &  35  & 0\farcs0025 \\
VLA    & 1996 Dec 31 &  1.4 &  12  & 1\farcs5 \\
VLA    & 1996 Dec 31 &  4.9 &  12  & 0\farcs43 \\
VLA    & 1996 Dec 31 &  8.5 & 192  & 0\farcs25 \\
VLA    & 1996 Dec 31 & 14.9 &  20  & 0\farcs14 \\
MERLIN & 1997 Nov 30 &  5.0 & 750  & 0\farcs50 \\

\enddata 
\end{deluxetable}

\begin{figure}
\plotone{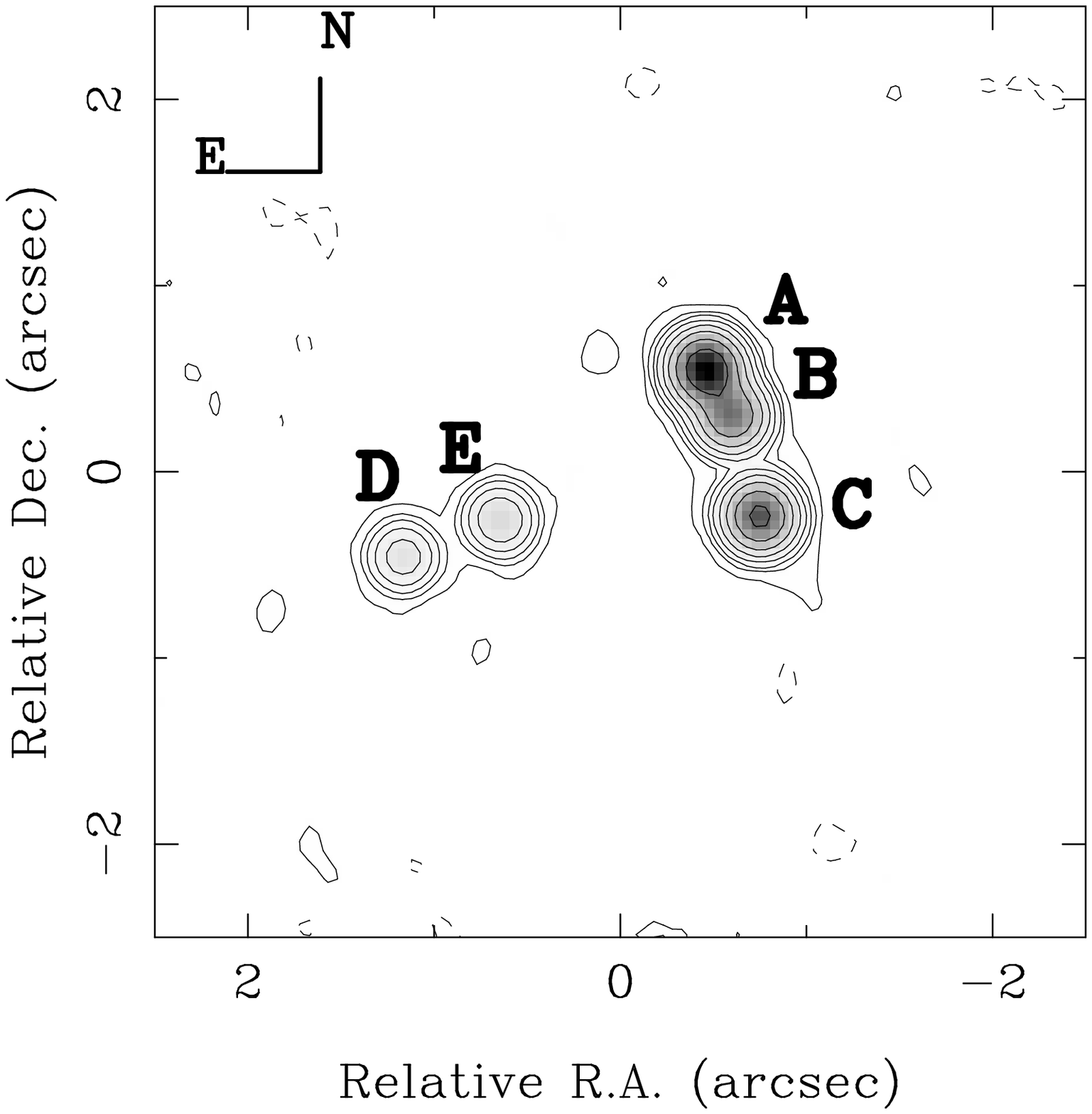}
\caption{\label{fig_vla_x} December 1996 8.5 GHz map.  The contours are 
($-$2.5, 2.5, 5, 10, 20, 40, 80, 160, 320, 640) times the RMS noise level
of 0.0145~mJy/beam.  Map made by fitting point source components,
with flux densities listed in Table~\ref{tab_radflux}, to the $(u,v)$ data,
and restoring with a
0\farcs27$\times$0\farcs23 restoring beam.}
\end{figure}
\begin{figure}
\plotone{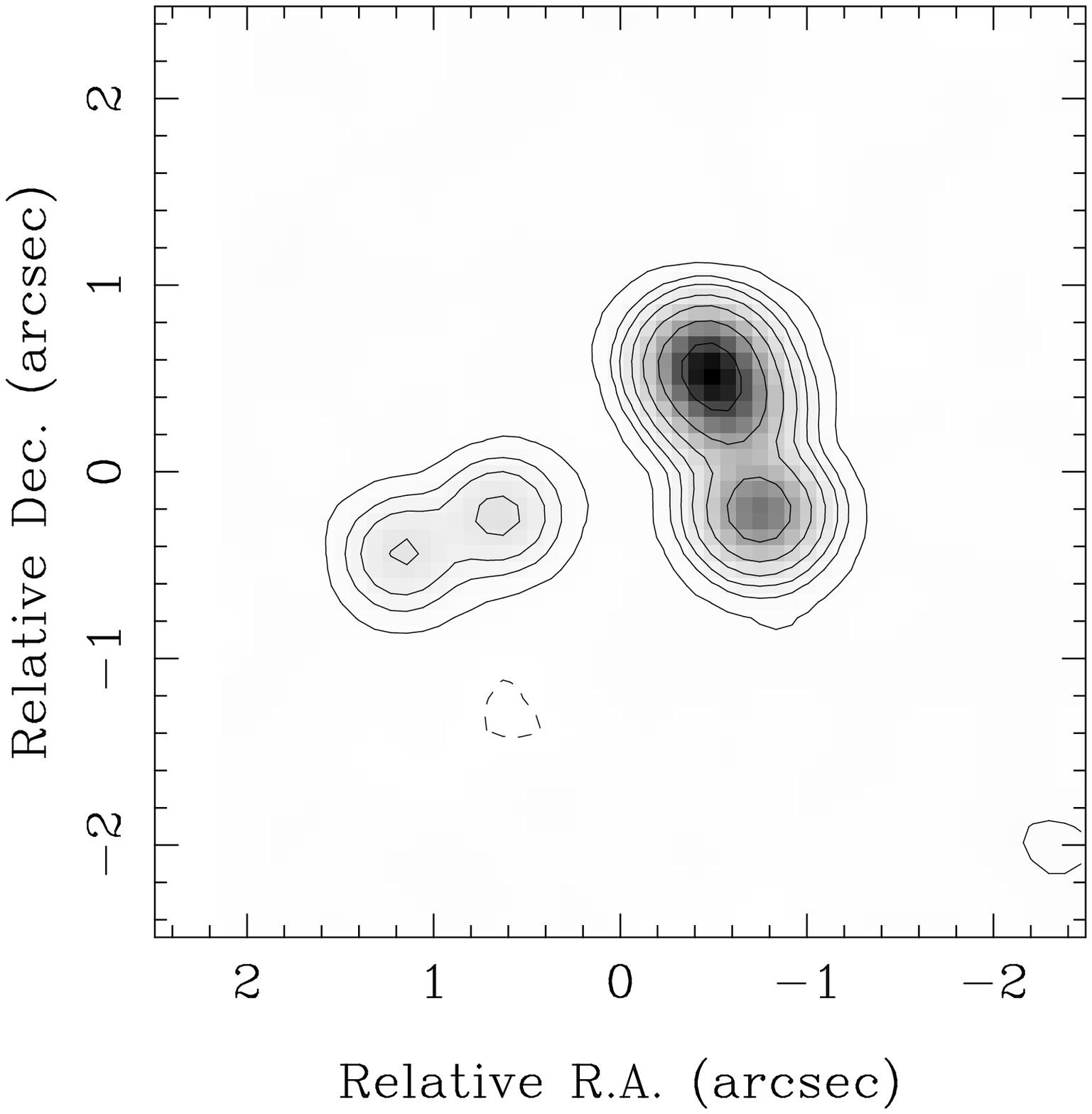}
\caption{\label{fig_vla_c} December 1996 4.9 GHz map.  The contours are 
($-$2.5, 2.5, 5, 10, 20, 40, 80, 160, 320) times the RMS noise level
of 0.0621~mJy/beam.  Map made by fitting point source components,
with flux densities listed in Table~\ref{tab_radflux}, to the $(u,v)$ data,
and restoring with a
0\farcs44$\times$0\farcs41 restoring beam.}
\end{figure}
\begin{figure}
\plotone{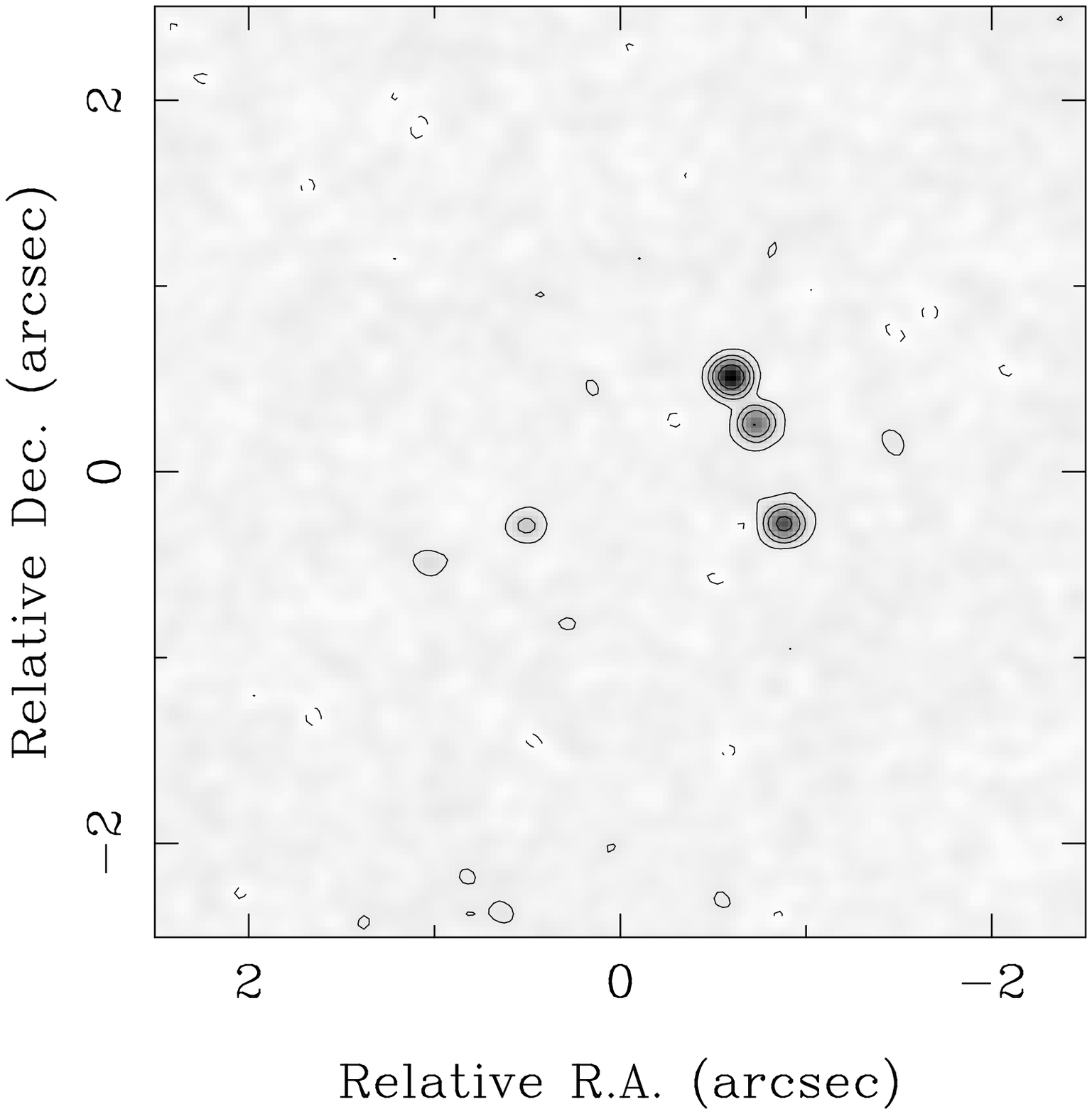}
\caption{\label{fig_vla_u} December 1996 15 GHz map.  The contours are 
($-$2.5, 2.5, 5, 10, 20, 40) times the RMS noise level
of 0.139~mJy/beam.  Map made by fitting point source components,
with flux densities listed in Table~\ref{tab_radflux}, to the $(u,v)$ data,
and restoring with a
0\farcs14$\times$0\farcs13 restoring beam.}
\end{figure}
\begin{figure}
\plotone{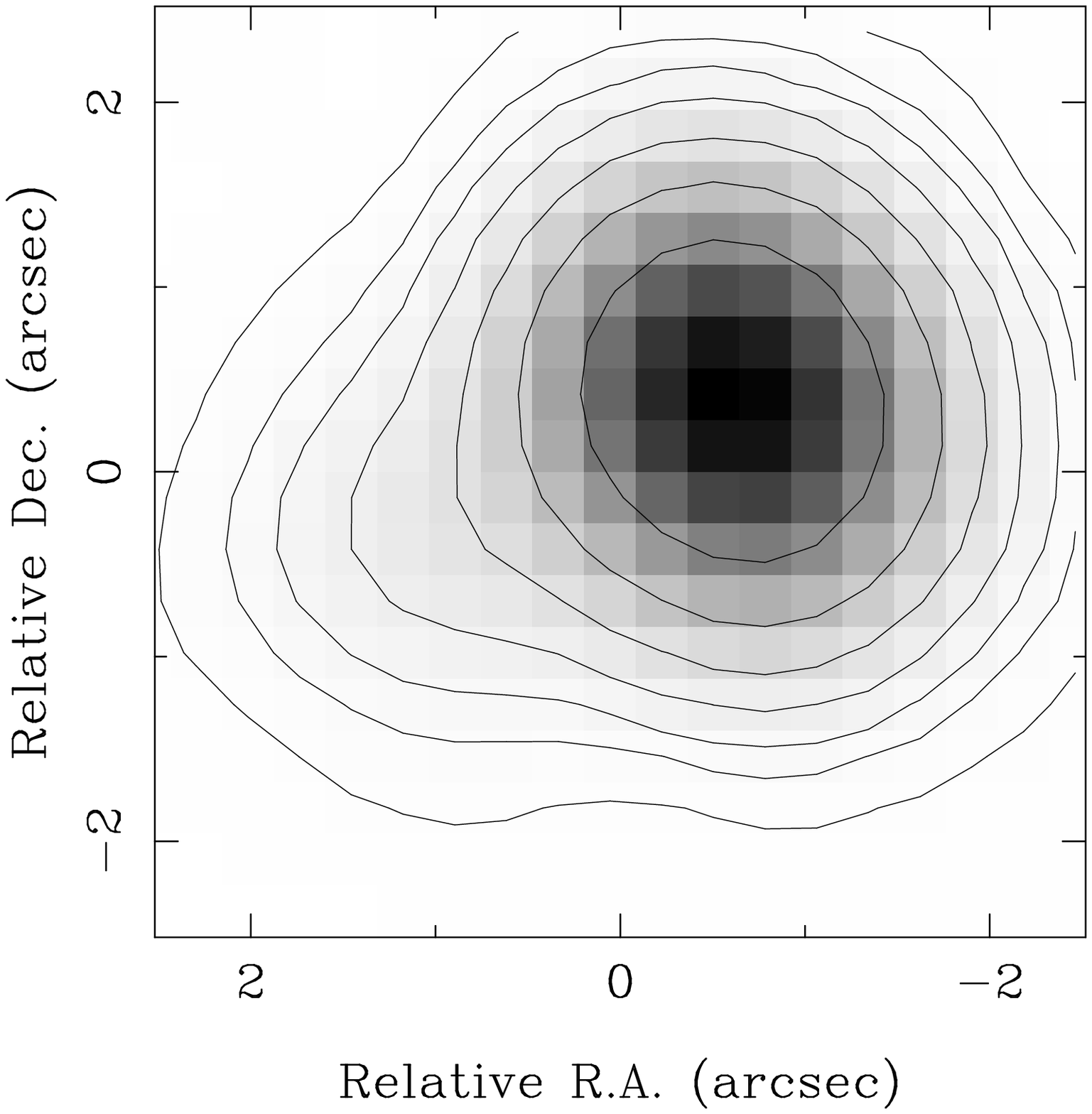}
\caption{\label{fig_vla_l} December 1996 1.4 GHz map.  The contours are 
($-$2.5, 2.5, 5, 10, 20, 40, 80, 160, 320) times the RMS noise level
of 0.0784~mJy/beam.  Map made by fitting point source components,
with flux densities listed in Table~\ref{tab_radflux}, to the $(u,v)$ data,
and restoring with a
1\farcs5$\times$1\farcs4 restoring beam.}
\end{figure}

The final maps are shown in Figures \ref{fig_vla_x} --
\ref{fig_vla_l}.  In all cases, the morphology seen in the discovery
map is duplicated, with five distinct components clearly present in
the high-dynamic-range 5 and 8.5~GHz maps.  Component positions and
flux densities were obtained using the model fitting procedures in
DIFMAP.  The brightest component (A) has an 8.5~GHz flux density of
16.55~mJy and is located at 20:47:20.29, +26:44:02.7 (J2000); the
positions of the other components relative to component A are given in
Table~\ref{tab_radpos} (see Figure \ref{fig_vla_x} for component
labels).  The positional uncertainties for the different components
are estimated as the beam size at each frequency divided by the
signal-to-noise ratio of the component flux densities; the positional
errors in the 8.5~GHz map are given in Table~\ref{tab_radpos}.  The
largest separation in the system is 1\farcs9.  At all frequencies,
five point components are adequate to fit the data.  No sign of
extended emission is seen, even in the deep 8.5~GHz map.  The final
component flux densities and RMS noise levels for the maps are given
in Table~{\ref{tab_radflux}}.  In all cases the noise levels are
within 5\% of the thermal noise expected in the maps.

The 8.5~GHz data were also analyzed to search for polarized
emission.  The polarization calibration was carried out in AIPS, using
3C~286 as the calibrator.  The phase calibrator 2115+295 was observed
over a range of parallactic angles and was used to determine the
instrumental polarization.  The final maps are consistent with
no polarized emission above 56$\mu$Jy/beam (0.34\% of the peak unpolarized
intensity).

\begin{figure}
\plotone{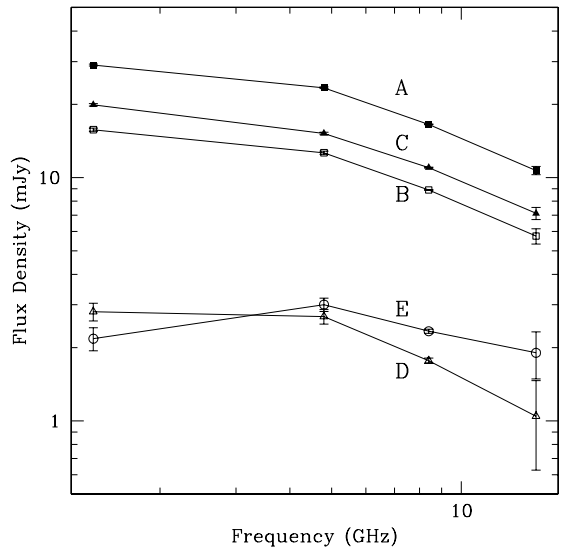}
\caption{\label{fig_radio_spec} Radio spectra of the five components in the 
B2045+265 system.  The error bars are 3 $\sigma$ errors, based on the RMS 
noise in the maps.}
\end{figure}

\begin{figure}
\plotone{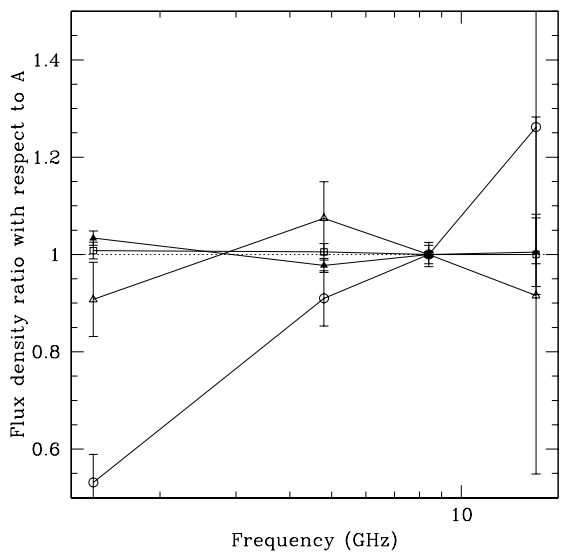}
\caption{\label{fig_flux_ratio} Flux density ratios for components B
(open squares), C (filled triangles), D (open triangles) and E (open
circles) with respect to component A.  All curves have been
normalized to their 8.5~GHz point.  Error bars are calculated by
propagating the RMS noise in the maps at each frequency.}
\end{figure}

Radio spectra for the five components are shown in
Fig. \ref{fig_radio_spec}.  Components A --
C have very similar spectra, with spectral indices of
$\alpha^{4.9}_{1.4} \sim -0.2$\, and $\alpha^{15}_{4.9} \sim -0.6$
($S_{\nu} \propto \nu^{\alpha}$).  The spectra of components D and E differ
from those of the three brighter components.  In order to determine
which, if either, of these two components is the counter image to
components A, B, and C, flux density ratios are computed for each
component with respect to component A.  By computing flux density
ratios at each frequency, it is possible to avoid uncertainties due to
errors in the absolute flux calibration.  The resulting curves (Fig.\
\ref{fig_flux_ratio}), normalized to their values at 8.5~GHz, clearly
indicate that the spectrum of component E differs from those of the
other four components.

\begin{deluxetable}{crrr}
\tablewidth{0pt}
\scriptsize
\tablecaption{\label{tab_radpos}Component Positions}
\tablehead{
\colhead{Component}
 & \colhead{$\Delta \alpha$\tablenotemark{a}}
 & \colhead{$\Delta \delta$\tablenotemark{a}}
 & \colhead{$\sigma_x$\tablenotemark{b}}
}

\startdata
A &   0.000  &   0.000  & \nodata \\
B & $-0.134$ & $-0.248$ & 0.001 \\
C & $-0.288$ & $-0.789$ & 0.001 \\
D &  +1.628  & $-1.007$ & 0.006 \\
E &  +1.121  & $-0.824$ & 0.005 \\

\enddata 
\tablenotetext{a}{Positions relative to component A are taken from
the 1996 Dec 31 8.5~GHz map.  Component A is at 20:47:20.29, +26:44:02.7 
(J2000)}
\tablenotetext{b}{Uncertainties in relative positions calculated using
the 1996 Dec 31 8.5~GHz data and assuming that the uncertainties in the 
component flux densities are three times the RMS noise level in the map.}
\end{deluxetable}

\begin{deluxetable}{llrllllll}
\tablewidth{0pt}
\scriptsize
\tablecaption{\label{tab_radflux}Component Flux Densities}
\tablehead{
\colhead{}
 & \colhead{}
 & \colhead{$\nu$}
 & \colhead{$S_A$}
 & \colhead{$S_B$}
 & \colhead{$S_C$}
 & \colhead{$S_D$}
 & \colhead{$S_E$}
 & \colhead{RMS} \\
\colhead{Date}
 & \colhead{Array}
 & \colhead{(GHz)}
 & \colhead{(mJy)}
 & \colhead{(mJy)}
 & \colhead{(mJy)}
 & \colhead{(mJy)}
 & \colhead{(mJy)}
 & \colhead{(mJy/beam)}
}

\startdata
1995 Sep 02
 & VLA
 & 8.5
 & 18.4
 & \phantom{0}9.42
 & 14.8
 &  2.41
 &  1.83
 &  0.28 \\
1995 Sep 11
 & VLA
 & 14.9
 & 15.5
 & \phantom{0}9.18
 & 11.8
 & 1.22
 & 2.36
 & 0.32 \\
1995 Nov 12
 & VLBA
 & 5.0
 & 15.8
 & \phantom{0}8.09
 & \phantom{0}8.75
 & \nodata
 & \nodata
 & 0.20 \\
1996 Dec 31
 & VLA
 & 1.4
 & 29.02 
 & 15.73
 & 19.92
 &  2.81
 &  2.18
 &  0.08 \\

 &
 & 4.9
 & 23.40
 & 12.65
 & 15.19
 &  2.68
 &  3.00
 &  0.06 \\

 &
 & 8.5
 & 16.55
 & \phantom{0}8.90
 & 10.99
 & 1.77
 & 2.34
 & 0.01 \\

 &
 & 14.9
 & 10.69
 & \phantom{0}5.75
 & \phantom{0}7.13
 & 1.05
 & 1.90
 & 0.14 \\
1997 Nov 30
 & MERLIN
 & 5.0
 & 16.8
 & \phantom{0}9.82
 & 14.8
 &  2.02
 &  1.53
 &  0.15 \\

\enddata 
\end{deluxetable}

\begin{deluxetable}{llllll}
\tablewidth{0pt}
\scriptsize
\tablecaption{\label{tab_fluxrats}Component Flux Density Ratios}
\tablehead{
\colhead{$\nu$ (GHz)}
 & \colhead{Date}
 & \colhead{B/A}
 & \colhead{C/A}
 & \colhead{D/A}
 & \colhead{E/A}
}

\startdata
\phantom{0}1.4 GHz
 & 1996 Dec 31
 & 0.54  $\pm$ 0.009
 & 0.69	 $\pm$ 0.010
 & 0.097 $\pm$ 0.008
 & 0.075 $\pm$ 0.008 \\
\phantom{0}4.9 GHz
 & 1996 Dec 31
 & 0.54 $\pm$ 0.009
 & 0.65	$\pm$ 0.009
 & 0.11\phantom{0} $\pm$ 0.008
 & 0.13\phantom{0} $\pm$ 0.008 \\
\phantom{0}8.5 GHz
 & 1995 Sep 02
 & 0.51 $\pm$ 0.051
 & 0.80 $\pm$ 0.059
 & 0.13\phantom{0} $\pm$ 0.046
 & 0.099 $\pm$ 0.046 \\
\phantom{0}8.5 GHz
 & 1996 Dec 31
 & 0.54 $\pm$ 0.003
 & 0.66 $\pm$ 0.003
 & 0.11\phantom{0} $\pm$ 0.003
 & 0.14\phantom{0} $\pm$ 0.003 \\
14.9 GHz
 & 1995 Sep 11
 & 0.59  $\pm$ 0.072
 & 0.76  $\pm$ 0.078
 & 0.079 $\pm$ 0.062
 & 0.15\phantom{0} $\pm$ 0.062 \\
14.9 GHz
 & 1996 Dec 31
 & 0.54  $\pm$ 0.044
 & 0.67  $\pm$ 0.047
 & 0.098 $\pm$ 0.046
 & 0.18\phantom{0} $\pm$ 0.063 \\

\enddata 
\end{deluxetable}

%
%
%

\section{MERLIN and VLBA Observations}

\begin{figure}
\plotone{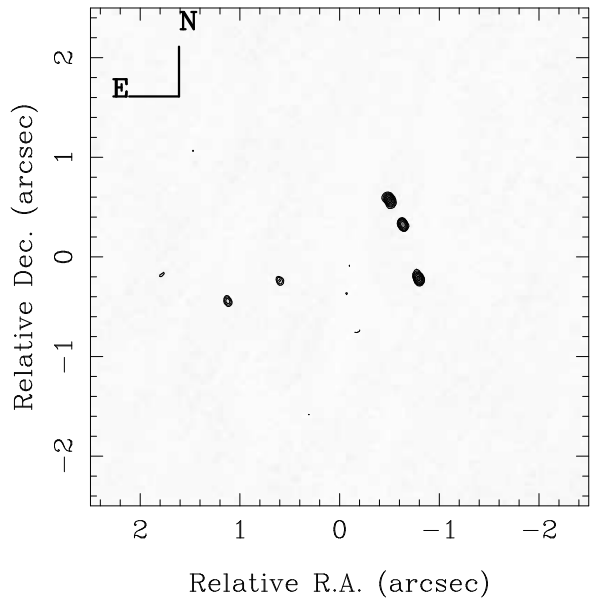}
\caption{\label{fig_merlin} MERLIN map.  Components A, B, and C may
be slightly resolved.}
\end{figure}

A 5~GHz MERLIN observation of B2045+265 was made on 1997 November 30,
with approximately 12.5~hr total integration on source.  The final map
(Fig.\ \ref{fig_merlin}) has an RMS noise level of 0.146~mJy/beam and
an angular resolution of 60~milliarcsec (mas).  The data are well
fitted by 5 point sources, with flux densities given in
Table~\ref{tab_radflux}.  However, there are some indications of
additional emission around components A and C in the residual map, at
the 2 -- 3\,$\sigma$ level.  The excess emission may indicate that the
images are slightly resolved.  If mas-scale structure can be detected
in the images, it can be used to put constraints on the lens model.
The transformation matrix between corresponding positions in the
resolved images provides crucial limits on the lensing potential by
fixing its second derivatives at those points (e.g., \cite{vlbi0957}).
Observations with high angular resolution and dynamic range are needed
to search for such structure.  A 5~GHz VLBA snapshot (35~min total
integration on source) of this system was made on 1995 November 12.
In this observation, which has an angular resolution of $\sim$1~mas,
the three brightest components are detected.  The flux density of each
component is only $\sim$60\% of its flux density in the 5~GHz VLA
observations.  Once again, this may indicate the presence of extended
mas-scale structure in this source which could be detected in more
sensitive observations.  Deep VLBA observations of B2045+265 have now
been scheduled which may produce maps with sufficient dynamic range to
detect extended structure in the images.

\section{Infrared Imaging}

\subsection{NIRC}

The system was imaged on 1996 July 31 using the Near Infrared Camera
(NIRC; \cite{nirc}) on the W.\ M.\ Keck I Telescope.  
Images were taken in both $J$ and $K$ bands, with 45 one-minute
exposures in $K$\ and 27 one-minute exposures in $J$.  In order to
estimate the point-spread function (PSF), 18 exposures of a star at a
distance of 1.47~arcmin were interleaved with the exposures on the
lens in each band.
The seeing was 0\farcs45--0\farcs90 in $K$ and 0\farcs75--1\farcs15 in
$J$\ during the observations.

\begin{figure}
\plotone{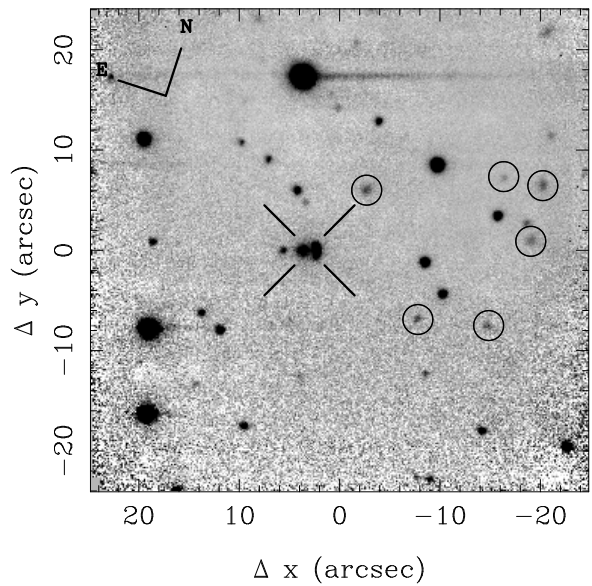}
\caption{\label{fig_k_mos} NIRC $K$ band mosaic of the B2045+265
field.  The lens system is marked with the crosshairs, with the
lensing galaxy to the left and an arc of emission from the background
source to the right.  The object 2\arcsec\ to the left of the galaxy
is a star.  The circled objects are extended and may be galaxies in
a group at the redshift of the lens (see \S~7.3).}
\end{figure}
\begin{figure}
\plotone{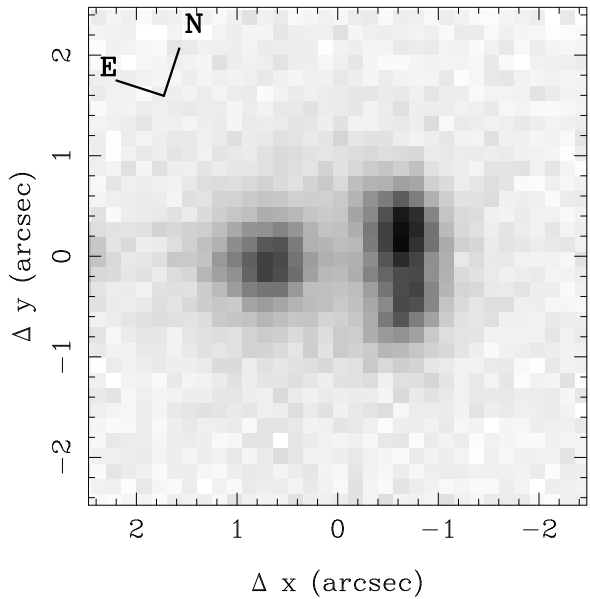}
\caption{\label{fig_nirc_k} NIRC $K$ band image of B2045+265.  The lensing
galaxy is to the left, perhaps with some emission being
contributed from component D.  An arc consisting of emission from components
A, B and C is to the right.}
\end{figure}
\begin{figure}
\plotone{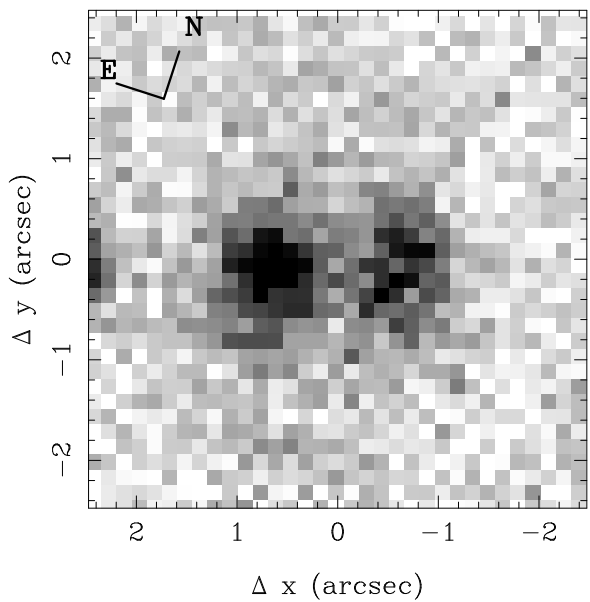}
\caption{\label{fig_nirc_j} NIRC $J$ band image of B2045+265.}
\end{figure}

The dark current level was subtracted from each image, and then
sky-subtraction and gain correction were performed.  The sky and gain
frames used for each image were constructed from images observed
directly before and after it.  The individual frames were aligned by
centroiding on a star which appeared in each frame.  For the highest
sensitivity in the final images, all 45 frames in $K$\ and all 27
frames in $J$\ were combined into mosaics.  The $K$ band mosaic is
shown in Fig.~\ref{fig_k_mos} and close-ups of the lens system in the
two bands are shown in Figs.~\ref{fig_nirc_k} and \ref{fig_nirc_j}.
The two main features in the images of the lens system are a short
arc, corresponding to the three brightest radio images, and the
lensing galaxy, which is located 1\farcs2 from the arc.  Spectroscopy
has shown that the object seen 2\arcsec\ to the left of the lens
system in Fig.~\ref{fig_k_mos} is a star.

\subsection{NICMOS\label{nicobs}}

\begin{figure}
\plotone{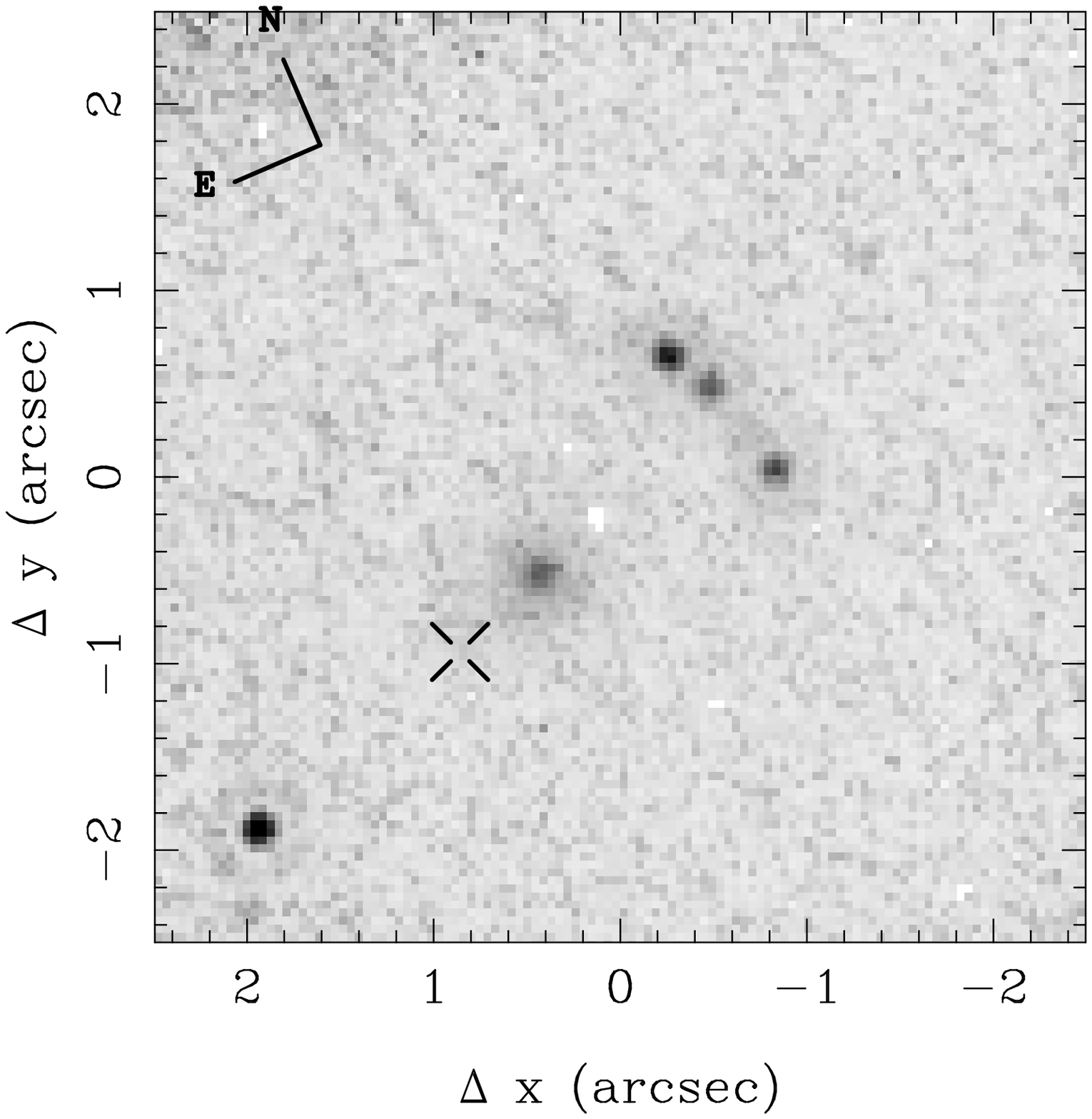}
\caption{\label{fig_nicmos} NICMOS F160W image of B2045+265.  The extended 
object near the center of the image is the lensing galaxy.  The 
crosshairs mark the expected position of component D.}
\end{figure}

B2045+265 was observed with the {\em Hubble Space Telescope} using
the Near Infrared Camera/Multi-Object Spectrometer (NICMOS) on 1997
July 14.  The NIC-1 camera was used, which has a pixel scale of
43~mas.  Two exposures were taken giving a total exposure time of
2624~sec. The images were subjected to the standard NICMOS calibration
pipeline involving bias and dark current subtraction, linearity and
flat-field correction, photometric calibration and cosmic ray
identification and removal. The final image (Fig.~\ref{fig_nicmos})
clearly shows the three brightest lensed images and the lensing
galaxy.  The relative positions of components A, B, and C match those
seen at radio wavelengths to within 0\farcs01.  In addition, the galaxy
location matches the position of radio component E to within the
errors.  There appears to be low surface brightness emission between
images B and C, indicating the possibility that extended optical
emission from the background source is being lensed into an arc-like
structure.  The expected position of component D is marked in the
figure.
 
\subsection{Photometry\label{photometry}}

Because the NICMOS image has high spatial resolution, it is possible
to compute magnitudes for the individual images of the background
source in the F160W bandpass (roughly corresponding to the
ground-based $H$ band).  Magnitudes were calculated using both the
DAOPHOT package (\cite{daophot}) in IRAF\footnote{IRAF (Image
Reduction and Analysis Facility) is distributed by the National
Optical Astronomy Observatories, which are operated by the Association
of Universities for Research in Astronomy under cooperative agreement
with the National Science Foundation.} and the SExtractor package
(\cite{SExtractor}).  First, the magnitudes of the three bright lensed
images were calculated using 0\farcs26 diameter apertures.  Next, an
empirical PSF, constructed from radial profiles of stars in the field,
was fitted to the emission from the lensed images and the star near
the lensing galaxy.  The scaled PSFs were then subtracted from the
data and the lens magnitude was calculated using a 1\farcs9 diameter
aperture.  This aperture was chosen to match the largest image
separation in the system, which is approximately equal to twice the
Einstein radius of the lens.  The aperture diameter corresponds to
7.9\,$h^{-1}$~kpc at the redshift of the lens.  The PHOTFNU header
card was used to convert count rates in the image into flux densities
into Janskys and then a Vega zero-point of 1087~Jy was assumed to get
the F160W magnitudes.  For comparison with the NIRC photometry (see
below), the magnitude of the ``arc'' defined by the three lensed
images of the background source was also computed.
The DAOPHOT and SExtractor packages produced magnitudes that were
consistent within the errors of the sky determination.  The final
magnitudes are given in Table~\ref{tab_irphot}.  It should be possible
to search for extinction caused by the lensing galaxy by comparing the
F160W and radio flux density ratios of the lensed images.  However,
the uncertainties in the NICMOS component magnitudes are large ($\sim
0.2^m$) and thus the F160W flux density ratios, B/A = 0.69 and C/A =
0.76, are consistent with the radio flux density ratios at the
1~$\sigma$ level.  The upper limit on the brightness of a
point-source at the location of component D is $m_{F160W} > 22^m$.
This is not surprising since component D should be $2.5^m$ fainter
than A.

It is considerably more difficult to do photometry on the NIRC images
because the seeing disk is large compared to the component
separations.  Instead of finding magnitudes of the individual lensed
images, the magnitude of the ``arc'' was calculated by centering a
rectangular aperture of width 0\farcs9 and height 1\farcs8 on the
``arc.''  The sky was estimated from regions directly to the north,
west and south of the aperture.  Once again, the DAOPHOT package was
used to subtract the emission from the lensed images and the star from
the data before the lens galaxy magnitude was calculated.  The
photometric zero points were estimated by observing the infrared
standard star SJ9184 ($K = 11.82$, $J = 12.18$; \cite{irstds}), which
was observed in each band prior to the lens and PSF star exposures.
The zero points in the two observed bands were $K_0 = 22.44,$ and $J_0
= 22.92$.  The $J$ and $K$ magnitudes of the lensing galaxy and
``arc'' are given in Table~\ref{tab_irphot}.

\begin{deluxetable}{crrccc}
\tablewidth{0pt}
\scriptsize
\tablecaption{\label{tab_irphot}Component Data (Infrared Observations)}
\tablehead{
 \colhead{Component}
 & \colhead{$\Delta \alpha$\tablenotemark{a}}
 & \colhead{$\Delta \delta$\tablenotemark{a}}
 & \colhead{$J$}
 & \colhead{$m_{F160W}$\tablenotemark{b}}
 & \colhead{$K$}
}

\startdata
A\tablenotemark{c}           & $ 0.00$ & $ 0.00$ & \nodata & 20.6 & \nodata \\
B\tablenotemark{c}           & $-0.13$ & $-0.23$ & \nodata & 21.0 & \nodata \\
C\tablenotemark{c}           & $-0.28$ & $-0.78$ & \nodata & 20.9 & \nodata \\
Lens Galaxy\tablenotemark{d} & $+1.10$ & $-0.80$ & 19.2    & 18.6 & 17.6    \\
``Arc''                      & \nodata & \nodata & 19.6    & 18.7 & 17.3    \\ 

\enddata 
\tablenotetext{a}{Relative positions in arcseconds calculated
with respect to component A after a $-23^{\circ}$ rotation was performed to
match the radio image orientation.}
\tablenotetext{b}{F160W magnitudes computed assuming that the Vega zero point 
is 1087~Jy.}
\tablenotetext{c}{Magnitudes computed in a 0\farcs26 diameter aperture.}
\tablenotetext{d}{Magnitudes computed in a 1\farcs9 diameter aperture.}
\end{deluxetable}


\section{Optical Spectroscopy\label{optspec}}

Spectra of B2045+265 were taken with the Low Resolution Imaging
Spectrograph (LRIS; \cite{lris}) on the Keck I Telescope on 1996 June
18--19.  A 1\arcsec\ longslit and the 300~gr/mm grating were used,
giving a pixel scale of 2.44~\AA/pix.  The slit was positioned at
P.A. = 112$^{\circ}$, in order to have the highest probability of
spatially separating the emission of the background source from that
of the lensing galaxy.  Seven 1500~sec exposures were taken, four on
the first night and three on the second night.

\begin{figure}
\plotone{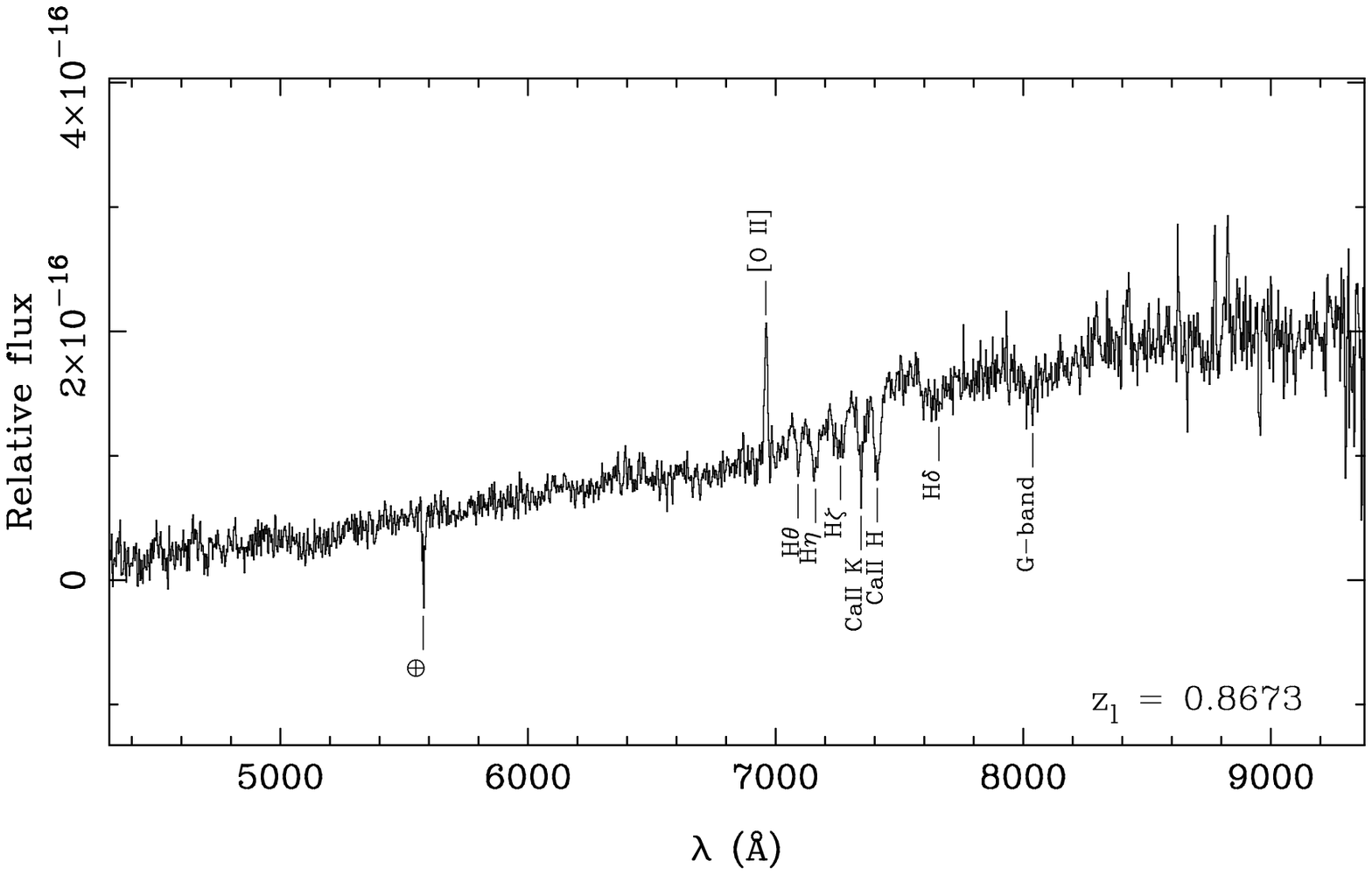}
\caption{\label{fig_spec_lens} LRIS spectrum of the lensing galaxy
from data taken on 1996 June 18--19.}
\end{figure}

\begin{figure}
\plotone{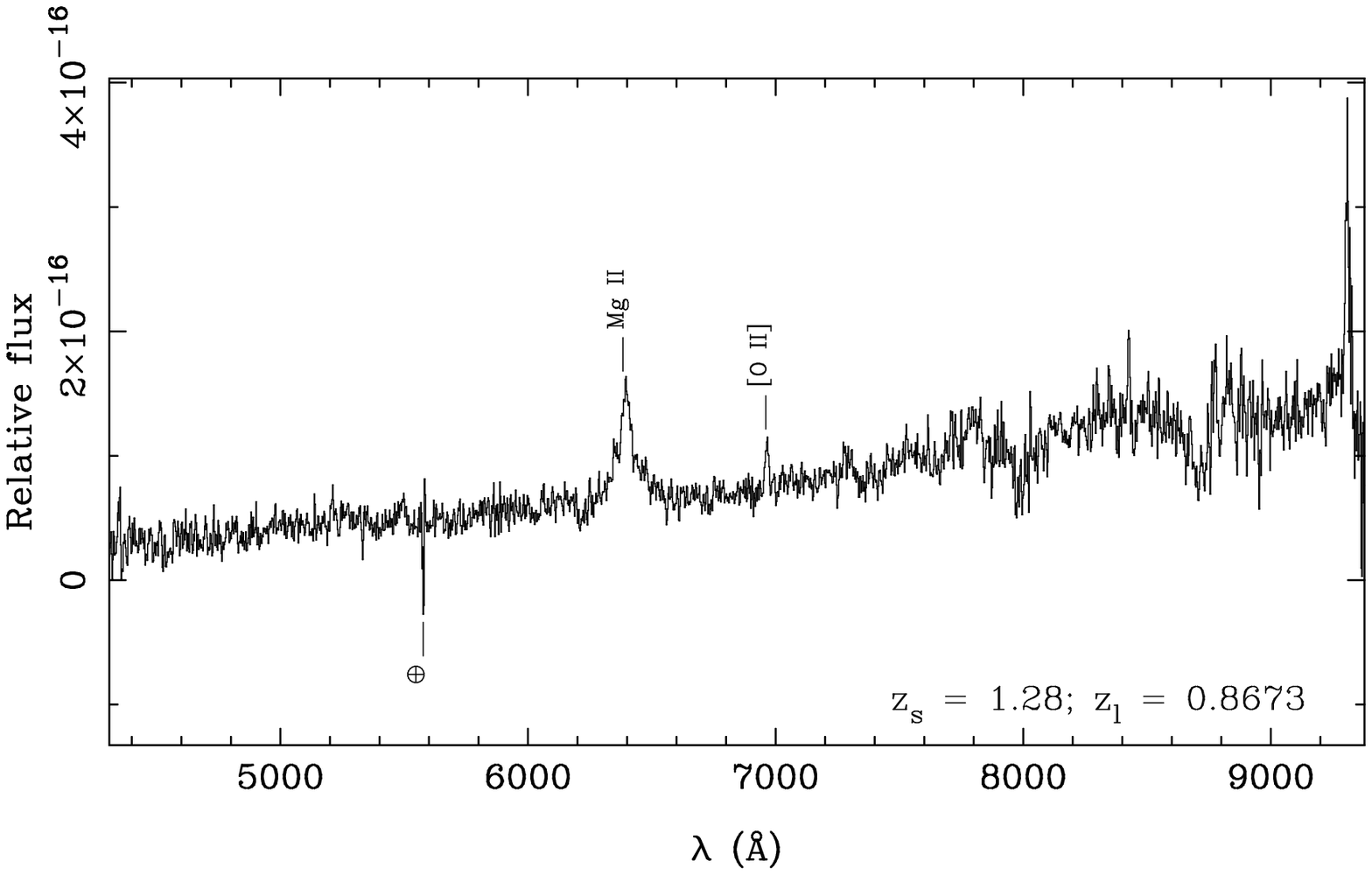}
\caption{\label{fig_spec_source} Spectrum of background source taken on
1996 June 18--19.  There is some contamination from light from the lensing
galaxy, which produces the [\ion{O}{2}] emission line.}
\end{figure}

The spectra were reduced using standard IRAF tasks.
The average seeing over the two nights was sufficiently good to enable
the extraction of separate spectra at the positions of the lensing
galaxy and the brightest source emission.  The spatial separation
between lens and source spectra is 1\farcs3, matching the separation
between the lens and the three bright lensed images seen in the NICMOS
image.
The final spectra have a wavelength range $\lambda\lambda$4309--9361\AA.  
Flux calibration was performed using the standard star Feige 110
(\cite{okestds}) which was observed at the end of each night. An
atmospheric absorption template was created by fitting a power law to
the flux-calibrated spectrum of the BL Lac object 2155--304.  This
template was used to remove atmospheric features from the B2045+265
spectra.  The individual exposures were weighted by the squares of their
signal-to-noise ratios and combined to create the final lens and
source spectra shown in Figs. \ref{fig_spec_lens} and
\ref{fig_spec_source}.

\begin{figure}
\plotone{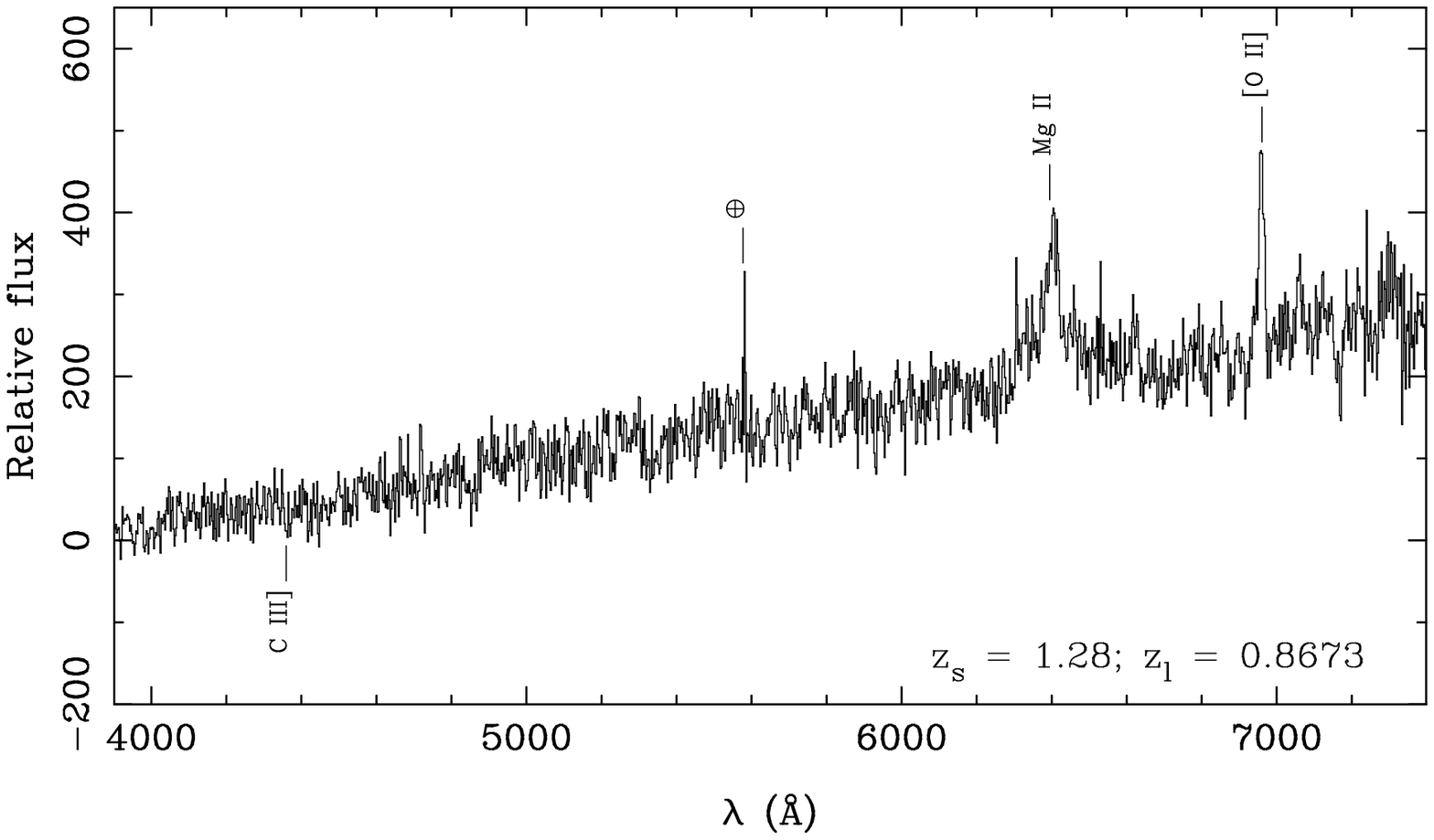}
\caption{\label{fig_spec_jlc} Wavelength calibrated spectrum taken on
1996 Sept.\ 15.  Both source and lens features are seen in the spectrum.
The position of the expected \ion{C}{3}] emission is marked.  The signal
drops to zero blueward of 4000\AA.}
\end{figure}

The lens system was re-observed with LRIS on 1996 September 15 with a
setup similar to that used in the earlier observations, but with
wavelength coverage $\lambda\lambda$2344--7398\AA.  Two 2000~sec
exposures were taken and the data were reduced using a method similar to
that described above.  In this case, however, 
the chip response was not removed from the final spectrum.  The seeing
was worse than in the earlier observations; as a result, the source and
lens spectra could not be extracted separately and both source and lens
features are seen in the final spectrum (Fig. \ref{fig_spec_jlc}).

The reduced spectrum of the lensing object is typical of a Sa galaxy
(e.g., \cite{galspecs}).  It shows [\ion{O}{2}] $\lambda$ 3727\AA\
emission at 6962\AA, \ion{Ca}{2} K $\lambda$3934\AA\ absorption at
7345\AA, \ion{Ca}{2} H $\lambda$3968\AA\ absorption (possibly blended
with H$\epsilon$ $\lambda$3970\AA) at 7411\AA, and H$\eta$\ and
H$\theta$ $\lambda\lambda$ 3835, 3797\AA\ absorption at 7159\AA\ and
7092\AA, respectively.  These spectral features establish the lens
redshift as $z_{\ell} = 0.8673 \pm 0.0005$, where the uncertainty in
the redshift is determined from the RMS scatter in the redshifts derived
from the [\ion{O}{2}], \ion{Ca}{2} K, H$\eta$\ and H$\theta$ lines.
The source spectrum shows a broad emission line at $\lambda =
6396$\AA.  No other broad emission features are seen in the spectrum,
which ranges from 4300 to nearly 9400\AA.  We identify the line as
\ion{Mg}{2} $\lambda$2800\AA, which implies $z_s = 1.28$, because it
is the only broad emission line that is so isolated in typical quasar
spectra (e.g., \cite{qspec1}; \cite{qspec2}).  The second set of
spectra were taken in order to search for \ion{C}{3}] $\lambda$1909
emission, expected to fall at $\sim 4350$\AA.  No line was seen at
this position, although the sensitivity in this part of the spectrum
is low and the presence of a weak line is not ruled out.  Other
possible identifications for the observed emission line
(e.g., H$\beta$, \ion{C}{3}], \ion{C}{4}, Ly$\alpha$) would imply the
presence of strong emission lines in parts of the spectrum in which
the signal-to-noise ratio is high, none of which we see.  We conclude that
the line is correctly identified and the source lies at redshift $z_s
= 1.28 \pm$0.01, where the uncertainty in the redshift is estimated from
uncertainties in finding the line centroid in both the emitted and the
observed frames.

\section{Source Variability}

In order for a lens system to be used to measure $H_0$, the background
source must be variable so that time delays can be measured.  We use
two methods to search for evidence of variability in the background
source in B2045+265.  The first is to examine the component flux
density ratios at different epochs.  The ratios are not affected by
errors in the absolute flux calibration and so any detected changes in
the flux density ratios reflect actual changes in the flux density of
the background source.  A comparison of the 8.5~GHz observations made
in September 1995 with those made in December 1996 shows small changes
in the ratios of the component flux densities, but not at a
significant level.

The second method is to monitor the total 22~GHz flux density of the
B2045+265 lens system.  Daily observations of the system were made
with the 40~m telescope of the Owens Valley Radio Observatory during
the periods 1996 May 18 -- September 03 and 1996 November 04 -- 22.
The data were calibrated using the CMBPROG package (\cite{emlthesis}).
A 15\% $\pm$ 3\%
change in total flux density was observed in November 1996 over a
period of less than a week.  Because several other sources being
monitored showed no change in flux density during the same period, the
observed change in flux density may indicate variability in the
background source.  However, it is also possible that the detected
variability is due to changes in component E, which has a radio
spectrum that is flatter that that of the lensed images.  To determine
which of the components is varying requires high angular resolution
monitoring of the system with the VLA.

\section{Discussion}

\subsection{The Nature of Component E\label{counterim}}

The detection of five unresolved radio components in the B2045+265
system raises the possibility that component E is the fifth lensed
image of the background source.  We believe, however, that component E
is instead associated with the lensing galaxy.  Although standard
lensing models can produce five images in configurations similar to
that seen in the B2045+265 system (e.g., \cite{bn}), the central fifth
image tends to be highly demagnified with respect to the other images.
This behavior is not seen in B2045+265; in fact, component E is
brighter than component D at high frequencies.  In addition, the radio
spectrum of component E (Figs.\ \ref{fig_radio_spec} and
\ref{fig_flux_ratio}) differs sufficiently from the spectra of the
other four components to suggest that component E is not related to
the other components.  Finally, the radio emission of component E is
spatially coincident with the infrared emission from the lensing
galaxy (see \S\ref{nicobs}).  We therefore conclude that component E
is indeed associated with the lensing galaxy.  It thus appears that
we have discovered a radio galaxy lensing a radio-loud quasar.  Of
previously known lens systems, only 2016+112 (\cite{2016_discovery})
may have radio emission associated with a lensing galaxy.  However,
recent observations of 2016+112 suggest that the radio emission of its
component C1 may be lensed emission from a (separate) background
source (\cite{2016_evn}).

The flat radio spectrum of component E in B2045+265 is typical of an
active galactic nucleus.  If component E is, indeed, associated with
the lensing galaxy, then the radio position of E gives an accurate
location for the nucleus of the lensing galaxy.  The location of the
center of the lensing galaxy relative to the lensed images is a key
component in models of the lensing potential (see \S\ref{model}).  An
{\em a priori} knowledge of the relative position of the center of the
lensing galaxy provides strong constraints on lens models.

\subsection{Preliminary Lens Model\label{model}}

We model the lens potential as a singular isothermal sphere
potential with ``mixed'' shear (\cite{csk91}), in which the scaled
lensing potential can be expressed as:
$$
\psi = b r + \gamma b r \cos 2(\theta - \theta_{\gamma}).
$$
Locations in the image plane relative to the lens center are given by
$\vec{x} = (r,\theta) = (x,y)$.  The critical radius of the isothermal
sphere is given by $b$, and $\gamma$\ and $\theta_{\gamma}$ give the
magnitude of the shear and its position angle.  In the modeling
process we have used an {\em astronomical} rather than mathematical
coordinate system such that $x \equiv \Delta \alpha$ and $y \equiv
\Delta \delta$.  The polar coordinates are defined in the usual way,
with $r = (x^2 + y^2)^{1/2}$ and $\theta = \arctan (y/x)$.  This
results in a {\em left-handed} coordinate system.

Although the image positions are defined relative to the
position of component A, they are treated as absolute positions.
Thus, both the lens position $(x_{\ell},y_{\ell})$\ and source
position $(x_s,y_s)$ are included as parameters in the model.  We
assume that component E marks the nucleus of the lensing galaxy and
fix the lens position at that location, leaving five varying
parameters and eight observational constraints ($x$\ and $y$\ for each
of the four images).  

\begin{figure}
\plotone{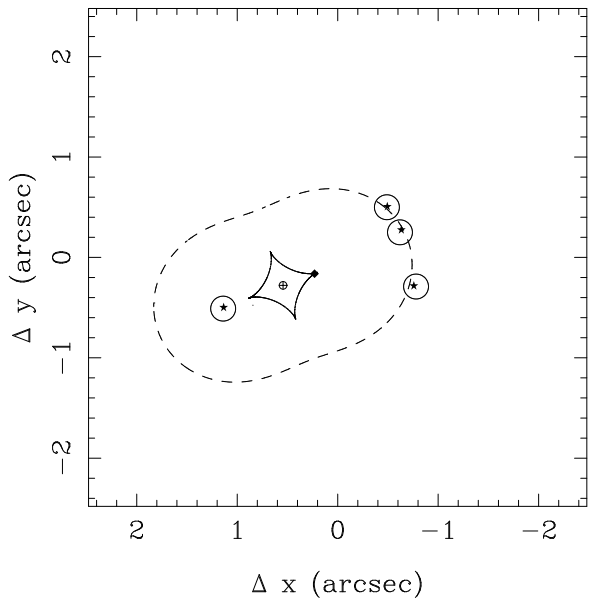}
\caption{\label{fig_model} Lens model showing critical curve (broken
line), caustic curve (solid line), lens center ($\oplus$), source position 
(filled diamond) and model image positions ($\star$).  The four circles are
centered on the observed image positions.}
\end{figure}

A downhill simplex routine (\cite{numrec}) is used to minimize the
difference between the observed and model positions, expressed as:
$$
\chi^2 = \sum_i \left[ \frac{|\vec{x}_m - \vec{x}_o|^2}{\sigma_x^2}\right]
$$
where the subscripts $m$\ and $o$ refer to the model and observed
values, respectively.  The positional errors are defined to be the
8.5~GHz beam size divided by the signal-to-noise ratio of the
component in the 8.5~GHz map.  To avoid biases introduced by the
choice of starting values, we repeat the process with a grid of
choices of initial conditions containing $3^n$ values, where $n$\ is
the number of varying parameters in the model fitting.  The best-fit
parameters for the model are given in Table~\ref{tab_models}, and the
model is shown graphically in Fig.\,\ref{fig_model}. The observed
image positions are recovered by this simple model of the lensing
potential, with RMS image displacements of $<$20~mas.

\begin{deluxetable}{lc}
\tablewidth{0pt}
\scriptsize
\tablecaption{Lens Model\label{tab_models}}
\tablehead{
   \colhead{Parameter}
 & \colhead{Value}
}
\startdata
$b$                                  & 1\farcs076   \\
$\gamma$                             & 0.104        \\
$\theta_{\gamma}$\tablenotemark{a}   & 69\fdg4      \\
$x_{\ell}$\tablenotemark{a}          & +1\farcs12\tablenotemark{b} \\
$y_{\ell}$\tablenotemark{a}          & $-$0\farcs82\tablenotemark{b} \\
$x_s$\tablenotemark{a}               & +0\farcs72   \\
$y_s$\tablenotemark{a}               & $-$0\farcs67 \\

\enddata
\tablenotetext{a}{The positions relative to component A ($x,y$) and 
$\theta_{\gamma}$ are defined in a {\em left-handed} coordinate system.  
See \S\ref{model}.}
\tablenotetext{b}{Held fixed during model fitting.}
\end{deluxetable}

Radio images are unaffected by microlensing, so it is possible to use
the flux density ratios of the images to constrain the model of the
lensing potential further.  The observed flux density ratios, however,
are difficult to reproduce with a simple model such as the
one presented above.  Although this simple model is adequate to
recover the four image positions, it does not reproduce the relative
magnifications of images A, B, C. The reason for this is easy to
see. If, as we have assumed, the potential is smooth, then we can
perform a Taylor expansion of it about the location of image B.  If we
approximate the images as collinear, then the highest order essential
terms in this expansion are quadratic in the coordinate perpendicular
to the A -- C line and quartic parallel to this line. Higher order
terms are, by assumption, ignorable. It is then possible to show that
in the limit of small image separations there are several scaling laws
including $S_B \sim S_A + S_C$, i.e., the flux of the central image
approaches the sum of the fluxes of the other two images
(\cite{kandp}; \cite{rdb_qjras}).  This is clearly violated in this
source, which indicates strongly that the potential is not smooth on
the scale of the A -- C image separation.

In order to improve upon our model, we have introduced higher order
terms in the potential such as might be produced by a local mass
perturbation. Specifically we expand the potential around images A and
C to third order along the A -- C line and to fourth order around image
B.  We then impose continuity of the potential and its first two
derivatives midway between A and B and midway between B and C. There
are then three observable quantities, the ratio of the image
separations and the ratios of the image fluxes, that can be used to
solve for the coefficients in these Taylor expansions. Performing this
exercise leads to the conclusion that adding a small, positive mass
perturbation centered on a point located 0\farcs11 from B towards C
suffices to recover the flux ratios as well as the image locations.
(This perturbation might be caused by a spiral arm, for example (cf
\cite{lenssub})). The ratio of the (almost certainly unmeasurable) time
delays $\Delta t_{BA}/\Delta t_{BC}$ is now found to be 0.2; however,
the delay $\Delta t_{BD}$ should be quite robust to this perturbation
within a given global model.  In conclusion, none of the existing
observations challenge the gravitational lens interpretation that we
present here.

\subsection{Properties of the Lensing Galaxy}

Properties of the lensing galaxy can be derived by using the fact that
it is acting as a gravitational lens.  For example, the mass within
the Einstein ring of a lens is (e.g., \cite{bn}):
$$
M_E \approx 1.24 \times 10^{11} 
\left( \frac{\theta_E}{{\rm 1\arcsec}} \right)^2
\left( \frac{D}{\rm 1 Gpc} \right ) M_{\sun}, 
\quad D \equiv \frac{D_{\ell} D_s}{D_{\ell s}}
$$ 
where $\theta_E$ is the angular radius of the Einstein ring and
$D_{\ell}$, $D_s$, and $D_{\ell s}$ are the angular diameter distances
between observer and lens, observer and source and lens and source,
respectively.  The angular diameter distance between two objects at
$z_i$ and $z_j$ ($z_i < z_j$) is given by (\cite{lambda}):
$$
D_{ij} = \frac{c}{H_0 (1+z_j)|\Omega_k|^{1/2}} 
{\rm sinn} \left ( |\Omega_k|^{1/2} {\cal I}(z_i,z_j) \right ),
$$
where
$$
{\cal I} (z_i,z_j) = 
\int_{z_i}^{z_j} 
\frac{dz}{[(1+z)^2(1+\Omega_M z) - z(2+z)\Omega_{\Lambda}]^{-1/2}}
$$
and ${\rm sinn}(x)$ is defined as $\sinh (x)$ for $\Omega_k<0$, $x$
for $\Omega_k=0$, or $\sin (x)$ for $\Omega_k>0$, respectively.  The
$\Omega$ terms are defined in the standard fashion:
$$
\Omega_M = \frac{8 \pi G}{3 H_0^2} \rho_0, \quad
\Omega_{\Lambda} = \frac{\Lambda}{3 H_0^2}, \quad
\Omega_k = -\frac{k c^2}{R_0^2 H_0^2}.
$$
The integral $\cal I$ has an analytic solution for $\Omega_{\Lambda} =
0$, and if $\Omega_M = 1$ the angular diameter distance acquires the
simple form
$$
D_{ij} =
\frac{2 c}{H_0} \frac{1}{(1+z_j)} 
\left( \frac{1}{(1+z_i)^{1/2}} - \frac{1}{(1+z_j)^{1/2}} \right ).
$$

For B2045+265, the lens redshift is secure.  The source redshift,
based on one emission line, is less certain.  For this reason, we
express the physical quantities derived in this section as functions
of the source redshift.  We will also assume, for ease of discussion,
an Einstein-de~Sitter cosmology.  The effects of varying the source
redshift and the cosmological model are presented in
Table~\ref{tab_lenspars}.  The angular diameter distances for this system
are $D_{\ell} = 860\,h^{-1}$Mpc and
$$
\frac{D_{s}}{D_{\ell s}} = 
\frac{1.8673[(1+z_s)^{1/2} - 1]}{(1+z_s)^{1/2} - 1.8673}.
$$
The mass enclosed within the Einstein ring radius is calculated by
assuming that the maximum image separation is a measure of
$2\theta_E$; i.e., that $\theta_E$=0\farcs95.  This assumption yields
$M_E \approx 9.6 \times 10^{10} (D_s / D_{\ell s}) h^{-1} M_{\odot}$
for the B2045+265 system.  Note that $M_E$ represents the {\em total}
mass contained in the cylinder with projected radius $R_E$.  The speed
of a particle moving in a circular orbit ($v_{circ}$) at the Einstein
ring radius is derived from $M_E$ by assuming that the mass
distribution of the lensing galaxy is an isothermal sphere.  A simple
calculation is used to convert $M_E$ to the mass inside the sphere of
radius $R_E$.  The value for $v_{circ}$ follows directly from the mass
inside the sphere.  The luminosity of the lens inside $R_E$ can be
estimated by converting its $K$ magnitude to rest-frame $M_B$ and
$M_V$ magnitudes.  The $k$-correction for the lens redshift and
rest-frame ($V - K$) and ($B - V$) colors for typical Sa galaxies are
taken from Poggianti (1997).  The resulting luminosities are $(L_V)_E
= 2.92 \times 10^{10} h^{-2} L_{\sun ,V}$ and $(L_B)_E = 2.36 \times
10^{10} h^{-2} L_{\sun ,B}$.  The rest-frame $B$ band mass-to-light
ratio becomes $(M/L_B)_E = 4.1 (D_s / D_{\ell s}) h (M/L_B)_{\sun}$.

Inserting the source redshift derived from the optical spectra
(\S\ref{optspec}) yields a lens mass of $M_E \approx 4.7 \times
10^{11} h^{-1} M_{\odot}$ within the Einstein ring radius ($\sim4
h^{-1}$~kpc at the redshift of the lens).  This mass implies a
circular velocity at $R_E$ of 570~km/sec, two to three times the
velocities seen in both nearby (e.g., \cite{sa_rotation}) and more
distant ($0.1 < z < 1$; \cite{vogt}) spiral galaxies.  The Einstein
ring mass-to-light ratios are $(M/L_V)_E = 16 h (M/L_V)_{\sun}$ and
$(M/L_B)_E = 20 h (M/L_B)_{\sun}$.  These mass-to-light ratios are not
unusual for lens systems (e.g., Fig.~7 of \cite{lensgal}), although
they are higher than expected given the redshift of the lens and
assumptions about the luminosity evolution of lens galaxies (Keeton et
al.\ 1998).  The large lensing mass required to produce the observed
image separation is a result of the relative closeness of the
background source to the lens in redshift space.  There are several
possible explanations for the unusually high projected mass.

Compact groups of galaxies associated with the primary lensing
galaxies have been discovered in recent observations of two lens
systems (1115+080 and B1422+231; Kundi\'{c} et al.\ 1997a,b;
\cite{t1115_1422}).  We are conducting observations of the B2045+265
field to search for evidence of a group associated with the lensing
galaxy (see also Fig.~\ref{fig_k_mos}).  The high contamination rate
by field galaxies ($\sim$85\% at $z \sim 1$; \cite{cluster_spec}),
however, makes high-redshift groups difficult to detect.  If such a
group exists and is gravitationally bound, its gravitational potential
can enhance the image splitting by the lensing galaxy.  In such a
scenario, we would overestimate the mass of the lensing galaxy.

Another explanation for the high projected mass inside the Einstein
ring is that the lens may be one of a close pair of possibly merging
galaxies.  This situation has been observed in the case of the CLASS
lens 1608+656.  The initial observations revealed an Einstein ring
projected mass of $M_E \approx 3.1 \times 10^{11} h^{-1} M_{\odot}$,
corresponding to $v_{circ} = 440\,{\rm km/sec}$ (\cite{m1608};
\cite{f1608}).  Later HST imaging showed that the lens is composed of
a very close pair of galaxies, explaining the high mass and
ellipticity required for the 1608+656 system (\cite{hstlens}).  We are
unable to determine unequivocally whether two galaxies are present in
the B2045+265 system because the sensitivity of our NICMOS observation
is too low.  More sensitive NICMOS and WFPC2 observations of B2045+265
are needed to test the two-galaxy hypothesis.

A third possibility is that the source redshift or the assumed
cosmology (or both) are in error.  We examine the results of varying
the source redshift by assigning different identifications to the
emission line seen in Fig.~\ref{fig_spec_source}.  We then
re-calculate the physical parameters associated with the lens in three
different cosmological models: $(\Omega_M, \Omega_\Lambda) =$ (1, 0),
(0.2, 0), and (0.2, 0.8).  We find that the effects of changing $z_s$
are larger than those produced by changing the cosmological model.  Of
the alternative identifications of the emission line in the source
spectrum, the choice of Ly$\alpha$ produces lens galaxy properties
that are the closest to those associated with normal nearby spirals
(Table~\ref{tab_lenspars}).  However, no sign of a Ly$\alpha$ forest
is seen on the short-wavelength side of the emission line, even after
subtracting the estimated contribution of the lens galaxy from the
source spectrum.  The absence of other emission lines in the spectrum
(\S\ref{optspec}) and the lack of a Ly$\alpha$ forest argue against
the identification of the emission line as Ly$\alpha$.  More sensitive
spectroscopy of this system must be conducted to determine the source
redshift unambiguously.

\begin{deluxetable}{cccccccccc}
\tablewidth{0pt}
\scriptsize
\tablecaption{Lens Parameters\label{tab_lenspars}}
\tablehead{
   \colhead{}
 & \colhead{}
 & \colhead{}
 & \colhead{}
 & \colhead{$D_{\ell}$}
 & \colhead{$D_s$}
 & \colhead{$D_{\ell s}$}
 & \colhead{$M_E$}
 & \colhead{$(M/L_B)_E$}
 & \colhead{$v_{circ}$} \\
   \colhead{Line ID}
 & \colhead{$z_s$}
 & \colhead{$\Omega_M$}
 & \colhead{$\Omega_\Lambda$}
 & \colhead{($h^{-1}$ Mpc)}
 & \colhead{($h^{-1}$ Mpc)}
 & \colhead{($h^{-1}$ Mpc)}
 & \colhead{($10^{11} h^{-1} M_\odot$)}
 & \colhead{($h (M/L_B)_\odot$)}
 & \colhead{(km/sec)}
}
\startdata
\ion{Mg}{2} & 1.28 & 1.0 & 0.0 & \phantom{1}860 & \phantom{1}890 & 180 & 4.7 & 20 & 570 \\
            &      & 0.2 & 0.0 &           1000 &           1100 & 240 & 5.3 & 16 & 560 \\
            &      & 0.2 & 0.8 &           1200 &           1300 & 340 & 5.1 & 11 & 500 \\
\ion{C}{3}] & 2.35 & 1.0 & 0.0 & \phantom{1}860 & \phantom{1}810 & 330 & 2.3 & 10 & 400 \\
            &      & 0.2 & 0.0 &           1000 &           1200 & 470 & 2.8 & \phantom{1}9 & 400 \\
            &      & 0.2 & 0.8 &           1200 &           1300 & 650 & 2.6 & \phantom{1}6 & 360 \\
\ion{C}{4}  & 3.13 & 1.0 & 0.0 & \phantom{1}860 & \phantom{1}740 & 350 & 2.0 & \phantom{1}8 & 370 \\
            &      & 0.2 & 0.0 &           1000 &           1100 & 530 & 2.5 & \phantom{1}8 & 380 \\
            &      & 0.2 & 0.8 &           1200 &           1200 & 700 & 2.3 & \phantom{1}5 & 340 \\
Ly $\alpha$ & 4.26 & 1.0 & 0.0 & \phantom{1}860 & \phantom{1}640 & 337 & 1.8 & \phantom{1}7 & 360 \\
            &      & 0.2 & 0.0 &           1000 &           1100 & 540 & 2.3 & \phantom{1}7 & 370 \\
            &      & 0.2 & 0.8 &           1200 &           1100 & 690 & 2.1 & \phantom{1}5 & 330 \\
\enddata
\end{deluxetable}


\subsection{Prospects for Measuring $H_0$}

A lens system can be used to estimate $H_0$ by comparing the time
delays predicted by the lens model to the observed time delays.  The
time delays are a function of the lensing potential, the
source and image positions, the redshifts of the source and lens,
and the world model (through the angular diameter distances):
$$
\Delta t_i = (1 + z_{\ell})\frac{D_{\ell} D_s}{c D_{\ell s}}
   \left[\frac{1}{2} |\vec{x}_i - \vec{\beta}|^2 - \psi(\vec{x}_i) \right],
$$
where $\vec{x}_i = (x_i,y_i)$ is the angular position of the $i^{th}$
lensed image and $\vec{\beta} = (x_s,y_s)$ is the position of the source.
The observable quantity is the difference in the time delays, $\Delta
t_{ij} \equiv \Delta t_i - \Delta t_j$.  The predicted delays are
proportional to $h^{-1}$ through the ratio of angular diameter
distances, so if it is possible to measure delays, the ratio of
the predicted and observed delays will give $h$.  For the model given
in \S\,\ref{model}, the components are expected to vary in the order:
C, A, B, D.  The predicted delays between components A, B and C are
small: $\Delta t_{AC} = 6.5 h^{-1}$\,hr and $\Delta t_{BC} = 7.7
h^{-1}$\,hr.  The delays between each of the three bright components
and component D are all on the order of 142\,$h^{-1}$\,d.  It will be
challenging to measure the short time delays since radio loud quasars
typically do not vary significantly on those time scales.  However,
with high-sensitivity observations, it will be quite possible to
measure the long delays in this system, if the background source is
variable.

\section{Summary and Future Work}

We have discovered a new four-image gravitational lens in the second
phase of CLASS.  The B2045+265 lens system is unusual in several
respects. First, it is, with the possible exception of the puzzling
2016+112 system, the first known gravitational lens system in which
both the background source and the lensing galaxy are radio sources.
The radio emission from the lens, presumably from an active nucleus,
can be used to locate the center of the lensing galaxy with high
precision, which provides a strong constraint for lens models.
Secondly, the flux density ratios of the three brightest lensed images
are inconsistent with the predictions of simple models of lensing
potentials.  This result may indicate that there is significant
structure in the mass distribution.  Finally, the projected mass (and
mass-to-light ratio) inside the Einstein ring of the lens is unusually
high for a Sa galaxy.  This may imply that there is a significant
amount of dark matter associated with this galaxy, that the lens
actually consists of a close pair of galaxies, or that the overall
mass distribution of an associated group of galaxies is contributing
to the lensing potential.  Alternatively, the source redshift may be
misidentified.

There are intriguing hints that the images of the background source
may be resolved by high angular resolution observations.  We have
obtained high dynamic range VLBA observations which can be used to
search for this possible extended mas-scale structure.  If such
structure is detected, the transformation matrices between the lensed
images of the background source provide strong constraints on the
lensing model.  These data may explain the unusual flux density ratios
among the three bright images.  The VLBA observations also will
provide a more accurate position for the center of the lensing galaxy.
In addition, more sensitive {\em Hubble Space Telescope} and
spectroscopic observations can provide further details about the
lensing galaxy and background source.  The goal of these observations
is to develop a well-constrained model of the lensing potential.  A
program of monitoring can be used to search for time delays in the
system.  Any measured delays can be combined with the lens models to
yield a measurement of $H_0$.

\acknowledgments 

We are indebted to the VLA analysts and operators, to Terry Stickel
and Wayne Wack for heroic operation of the Keck Telescopes, and to the
Keck Observatory and STScI staff.  CDF thanks Erik Leitch for
generously spending many hours in instruction in the use of the OVRO
40~m telescope and for writing the CMBPROG software.  The operation of
the 40~m telescope would be impossible without the dedication and
knowledge of Russ Keeney and Mark Hodges.  We are grateful to Gerry
Neugebauer, Lee Armus and Aaron Evans for their expert assistance
during the reduction of the NIRC and NICMOS data.  We thank Lori
Lubin, D. Wardell Hogg, Mike Pahre, Mark Metzger, and Chung-Pei Ma for
useful discussions and comments on the manuscript.  We thank the
anonymous referee for helpful suggestions on how to improve the paper.
This work is supported by the NSF under grant \#AST 9420018 and by the
European Commission, TMR Program, Research Network Contract
ERBFMRXCT96-0034 ``CERES.''

\end{document}